\documentclass[fleqn,usenatbib]{mnras}

\usepackage{newtxtext,newtxmath}

\usepackage[T1]{fontenc}
\usepackage{ae,aecompl}

\usepackage{graphicx}	
\usepackage{amsmath}	
\usepackage{amssymb}	

\newcommand{\hi}{H{\sc i} \,}
\newcommand{\hii}{H{\sc i}\,21cm}
\newcommand{\Ms}{\textrm{M}_\odot}
\newcommand{\gasfrac}{\textrm{M}_{\textrm{HI}}/\textrm{M}_*}
\defcitealias{NatPaper}{vD18}

\title[\hi in NGC1052-DF2]{A dearth of atomic hydrogen in NGC1052-DF2}

\author[A. Chowdhury]{
Aditya Chowdhury$^{1}$\thanks{E-mail: chowdhury@ncra.tifr.res.in}\\
$^{1}$National Centre for Radio Astrophysics, Tata Institute of Fundamental Research (NCRA-TIFR), Pune 411007, India. 
}

\date{Accepted: 2018 October 8. Received: 2018 September 28; in original form: 2018 August 31}

\pubyear{2018}

\begin{document}
\label{firstpage}
\pagerange{\pageref{firstpage}--\pageref{lastpage}}
\maketitle

\begin{abstract}
The recently claimed discovery of an ultra-diffuse galaxy lacking dark matter has important implications for alternate theories to dark matter as well as models of galaxy formation in the lambda cold dark matter context. In this letter, we present a deep Giant Metrewave Radio Telescope search for atomic hydrogen in this ultra-diffuse galaxy, NGC1052-DF2. We report a non-detection of the \hii \ transition from the galaxy and place a stringent upper limit on the \hi mass of the galaxy -  $\textrm{M}_\textrm{HI} < 3.15 \times 10^6 \ ({\Delta V}/20 \ \textrm{km/s})^{1/2} \  \Ms \ (3\sigma)$. This makes NGC1052-DF2 an extremely gas-poor galaxy with an atomic gas to stellar mass fraction of $\textrm{M}_\textrm{HI}/\textrm{M}_* \ < \ 0.016 \ ({\Delta V}/20 \ \textrm{km/s})^{1/2} \ (3\sigma)$. Such low gas fractions are typical of dwarf ellipticals in dense environments and would be consistent with NGC1052-DF2 having undergone a tidal stripping event which can also explain its apparent lack of dark matter.
\end{abstract}




\section{Introduction}
Ultra Diffuse Galaxies (UDGs) are a class of low surface brightness galaxies with masses comparable to those found in dwarfs ($\textrm{M}_* \sim 10^8 \Ms$) but sizes of typical $\textrm{L}_*$ galaxies like the Milky Way ($\textrm{R}_e\sim 1.5-5$ kpc) \citep{vD15}. The initial discovery of such galaxies in cluster and group  environments  \citep[e.g][]{koda15,Martnez16,Merritt16,vanderBurg16,Yagi16} was seen as evidence for UDGs being failed $L_*$ galaxies, residing in large dark matter halos with early quenching of star formation \citep[for example][ found evidence for a UDG to reside in a milky way sized dark matter halo]{Dokkum16}. The quenching of these galaxies are attributed to gas stripping in high-density environment \cite[e.g.][]{Burkert17,Baushev18}. On the other hand, there have been observational evidences to suggest that contrary to the failed $L_*$ galaxy picture, most UDGs reside in dwarf-sized halos \citep[e.g][]{Beasley16,Amorisco18}. \citet{DiCintio17} used high-resolution N-body simulations to suggest that UDGs form in dwarf halos as a consequence of episodes of gas outflows driven by stellar feedback. The other possible formation scenario, suggested by \citet{Amorisco16}, is that UDGs are dwarf galaxies that reside in halos at the high end of the spin distribution leading to an extended disk. The discovery of $\sim 100$ isolated UDGs from a \hi selected sample by \citet{Leisman17} provided important clues to their formation. The authors found that UDGs do reside in dwarf-like halos as well as have a higher spin.

A recent analysis of the UDG NGC1052-DF2, a satellite of the elliptical galaxy NGC1052, by \citet{NatPaper} (hereafter \citetalias{NatPaper}) found evidence for anomalous kinematics which were interpreted as the galaxy containing very little dark matter.  The discovery, if confirmed, may have far-reaching implications for alternate theories of dark matter such as MOND and emergent gravity which predict that signatures of "dark matter" will always be detected in galaxies. \citet{Famaey18,Kroupa2018} investigate the dynamics of NGC1052-DF2 in the MOND paradigm and conclude that the current data are insufficient to rule out a MOND-based interpretation. On the other hand, the discovery is also interesting in the light of galaxy formation in the standard $\Lambda$CDM cosmology; for example, \cite{Ogiya18} found that tidal striping of NGC1052-DF2's progenitor in the gravitational field of the elliptical, NGC1052, can produce a galaxy with properties similar to those observed. 

The conclusions of \citetalias{NatPaper} depend primarily on two observational quantities - (i) The distance to NGC1052-DF2, and (ii) The velocity dispersion of the galaxy. \citet{Trujillo2018} found a distance to NGC1052-DF2 of 13 Mpc , placing the galaxy significantly closer  compared to the 20 Mpc distance found by \citetalias{NatPaper}. This revised distance estimate lead to a significantly reduced stellar mass and hence enough room for dark matter in the galaxy. Subsequently,  \citet{Dokkum2018} refuted this claim as a misidentification of the galaxy's Tip of the Red Giant Branch. Recently, \citet{Blakeslee2018} presented an independent analysis of the distance to NGC1052-DF2 supporting the original findings of \citetalias{NatPaper}.

The intrinsic velocity dispersion of the galaxy was determined by fitting the observed $\lambda$8,664.5 Ca{\sc II} \ lines to the spectra of 10 globular cluster like objects in the galaxy \citepalias{NatPaper}. Based on the biweight dispersion of the globular cluster velocities, these authors reported an intrinsic velocity dispersion of $\sigma_{\textrm{int}}=3.2^{+5.5}_{-3.2}$ and a 90$\%$ confidence upper limit of 10.5 km/s. \citet{Martin2018} argued that the biweight dispersion may be unsuitable for measuring velocity dispersion from such a small number of globular clusters. These authors used a Monte Carlo based analysis and found a 90\% confidence upper limit of 18.8 km/s on the intrinsic velocity dispersion of NGC1052-DF2. Their measurement implies a much larger dynamical mass, refuting the claim that the galaxy has little dark matter for its stellar mass. {\citet{Laporte18} also performed an independent analysis of the intrinsic dispersion of the galaxy based on the available globular cluster data and found a 95\% confidence upper limit of 20.5 km/s. Further, \citet{Hayashi18} found that the dynamical mass estimate depends on the assumed model of the globular cluster distribution and the low number of detected globular clusters when fitted with a Sersic profile leaves enough room for the galaxy to host a dark matter halo.}

An accurate measurement of the intrinsic velocity dispersion of the NGC1052-DF2 is thus necessary to confirm \citetalias{NatPaper}'s claim that the galaxy contains very little dark matter. One of the ways of doing so is by detecting the diffuse \hii \ emission from the galaxy. In this letter, we present deep \hii \ observations of NGC1052-DF2 with the Giant Metrewave Radio Telescope (GMRT). The observation and the analysis of data are discussed in section \ref{sec:obs}. In section \ref{sec:results}, we discuss the results and summarize this letter.

\section{Observations and Data Analysis}
\label{sec:obs}
We used the GMRT to observe NGC1052-DF2 (RA=02$^\textrm{h}$41$^\textrm{m}$46.8$^\textrm{s}$ DEC=$-08^\circ 24'12''$) for a total on-source time of $\sim 10$ h in 2018 June. We used the GMRT Wideband Backend in its 12.5 MHz bandwidth mode around 1411.75 MHz with 4096 channels. The spectral configuration allowed us to cover the entire velocity range of the NGC1052 group with a high velocity resolution of $\approx 0.64 $ km/s. NGC1052-DF2 is centred at a  velocity of $v_0=1802 \pm 2$ km/s \citepalias{NatPaper}, corresponding to a redshifted \hii \ centre frequency of 1411.914 MHz. For the purpose of this letter, we analyzed 512 channels around the line centre covering 1640.37 to 1964.15 km/s. We observed 3C48 to calibrate the flux scale and 0204-170 to calibrate the amplitude, phase and bandpass. 

The data were flagged for RFI using AOFlagger \citep{offringa-2012-morph-rfi-algorithm} and all subsequent flagging and analysis were done in CASA v5.3.0 \citep{CasaRef}. Standard data analysis procedures were followed to obtain phase, amplitude, and bandpass solutions on the phase calibrator. These amplitude solutions were scaled using the observations on 3C28 and transferred to the source. The source field is dominated by the central galaxy of the group, NGC1052, which hosts an AGN and has a NVSS reported L-band flux of 912 mJy \citep{Condon98}. NGC1052 is 13.5 arcmin away from the pointing centre and thus lies outside the GMRT primary beam half power point at these frequencies (at 1412 MHz, the primary beam half width at half maxima is 11.5 arcmin). Such a strong continuum source beyond the half power point made deconvolution using CLEAN very difficult due to systematic effects such as anisotropy of the primary beam, varying pointing errors as a function of time, etc. As a result of these difficulties, we self-calibrated the data using the following procedure : the CASA task {\sc fixvis} was used to bring NGC1052 at the phase centre and then, given that NGC1052 is at-least 10 times brighter than all other sources in the field, a standard point-source phase only calibration was done using the CASA task {\sc gaincal}. An amplitude selfcal was not done to avoid introducing errors in the flux scale from the time varying intensity of  NGC1052 as seen by the array due to effects such as primary beam rotation. A first-order polynomial fit was done on each continuum visibility and subtracted out using the CASA task {\sc uvcontsub}. Another phase rotation was performed on these continuum-subtracted visibilites to bring the target source back at the phase centre. 

NGC1052-DF2 has an effective optical radius of  $R_e=22.6$ arcsec \citepalias{NatPaper} and thus only  short baselines of the GMRT were used to ensure maximum line sensitivity. The spectral imaging was done with natural weighting and a long baseline uvcut of $4k\lambda$ ($\approx$ 1km) to produce spectral cubes with a spatial resolution of $43 \ \textrm{arcsec} \times 35 \ \textrm{arcsec}$. This cube was further convolved to a resolution of $1$ arcmin, ensuring that our measurements include most of the \hii \ emission from the galaxy.

\section{Discussion}
\label{sec:results}
\begin{figure}
    \centering
    \includegraphics[width=\columnwidth]{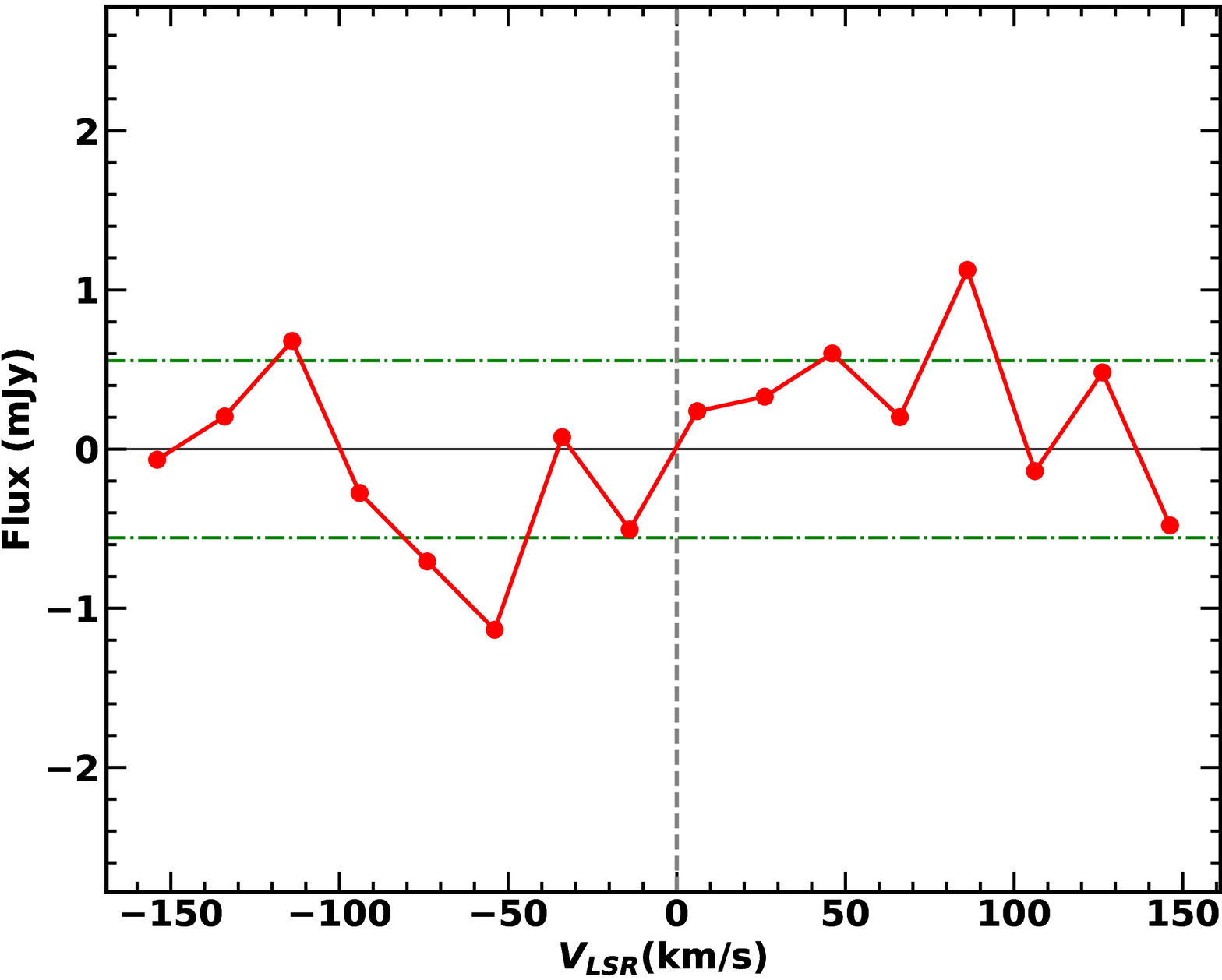}
    \caption{The spectra on NGC1052-DF2 around the redshifted \hii~line frequency at a velocity resolution of 20 km/s. The green dashed lines mark the $1\sigma$ noise of 0.56 mJy. The $3\sigma$ upper limit on the \hi mass of NGC1052-DF2 from this spectra is $3.15 \times 10^6 \ ({\Delta V}/20 \ \textrm{km/s})^{1/2} \  \Ms $.}
    \label{fig:spectra}
\end{figure}
\begin{figure}
    \centering
    \includegraphics[width=\columnwidth]{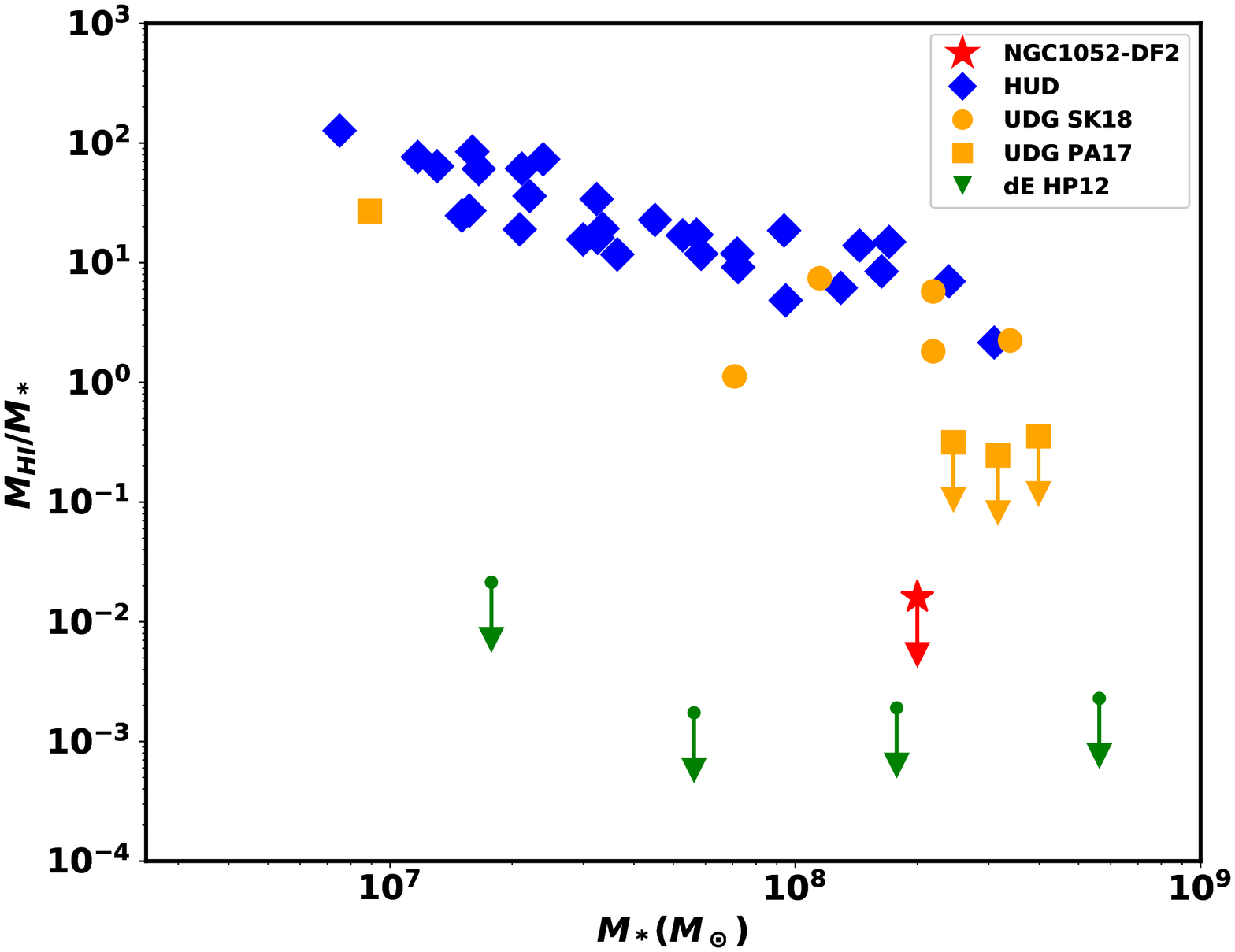}
    \caption{Comparision of the gas fraction of NGC1052-DF2 (red star in the plot) with other galaxies. The blue diamonds are the HUD-BG sample from \citep{Leisman17}. The orange squares are a sample of isolated UDGs from \citet[UDG PA17]{Papastergis17}.  The orange circles are a sample of blue UDGs around the Hickson Compact Groups from \citet[UDG SK18]{Spekkens18}. The green arrows are upper limits to gas-poor dwarf ellipticals in the Virgo cluster from \citet{Hallenbeck12}.}
    \label{fig:gasfrac}
\end{figure}
We searched for \hii \ emission from NGC1052-DF2 in the spectral cube at a wide range of velocity resolutions from 1.25 km/s to 30 km/s. These resolutions cover the wide range of reported uncertainties in the velocity dispersion of NGC1052-DF2(\citetalias{NatPaper},\citet{Martin2018}).  We did not detect the \hii \ line from the galaxy in any of the velocity resolutions. The spectra at each resolution was tested for gaussianity using the Anderson-Darling test and was found to be consistent with random sampling from a normal distribution. Figure \ref{fig:spectra} shows the spectra at a velocity resolution of 20 km/s resolutions. This velocity resolution is at the lower end of typical line widths of UDGs in  \citet{Leisman17}'s sample and is comparable to the 18.5 km/s 90 \% upper limit to the intrinsic  dispersion by \citet{Martin2018}. 

The rms flux density of the spectra shown in Fig. \ref{fig:spectra} is 0.56 mJy. {This flux density rms can be converted to a corresponding $3\sigma$ upper limit via the standard relation between the \hii \ line flux (S) and the \hi mass of the galaxy ($\rm M_{HI}$) : ${\rm M_{HI}/\rm M_\odot} = 2.35 \times 10^5  {\rm \langle D/Mpc \rangle}^2 
(\int {\rm S \ dV } \ / \ \rm Jy\,km/s)$, where D is the distance to the galaxy.}  With an assumed distance of 20 Mpc to NGC1052-DF2 \citep{Blakeslee2018,NatPaper,Dokkum2018}, we derive a $3\sigma$ upper limit to the \hi mass of NGC1052-DF2 of  $3.15 \times 10^6 \ ({\Delta V}/20 \ \textrm{km/s})^{1/2} \  \Ms $. For the stellar mass, $\textrm{M}_*=2\times 10^8 \Ms$ \citepalias{NatPaper}, we find a stringent upper limit on the atomic gas fraction of  $\gasfrac < 0.016 \ ({\Delta V}/20 \ \textrm{km/s})^{1/2} \ (3\sigma)$. {We note that this upper limit on the gas fraction is independent of the assumed distance to NGC1052-DF2 because both the stellar as well as the \hi mass has the same D$^2$ dependence.} 

Figure \ref{fig:gasfrac} compares the atomic gas fraction of NGC1052-DF2 to other galaxies.  We note that all reported detections of \hi from UDGs are in low-density environments \citep[e.g.][]{Papastergis17,Spekkens18}. For optically selected studies of UDGs, \citet{Spekkens18} found blue UDGs around the Hickson Compact Groups to be broadly gas rich. On the other hand, \citet{Papastergis17} found a mix population with three out of the four isolated UDGs targeted for \hi studies remaining undetected down to a gas fraction $\sim 0.5$.  These authors suggest that there is a dichotomy in the UDG population with the formation mechanism of gas-poor UDGs being poorly understood.  \citet{Leisman17} found a high mean gas fraction, $\gasfrac=35$, in their sample of isolated UDGs but their sample is \hi selected and thus is biased towards detecting galaxies with high \hi content. Overall, compared to most UDGs detected in low-density environments, we find  NGC1052-DF2 to be extremely gas-poor. The gas fraction upper limit is comparable to what is found for gas-poor dwarf ellipticals in cluster environments. For example, \citet{Hallenbeck12} found an upper limit of $\textrm{M}_\textrm{HI} < 10^{5.5} M_\odot (5\sigma)$  in a sample of dwarf  ellipticals in the Virgo cluster. The lack of atomic gas in NGC1052-DF2 suggests that either (a) UDGs in clusters and groups form via a different evolutionary path than their low-density counterparts \citep[see][for a formation mechanism of UDGs in clusters via tidal stripping and heating]{Carleton18} (b) NGC1052-DF2 is a special case of a UDG that underwent tidal stripping leaving the galaxy with little dark matter as well as gas \citep{Ogiya18}. 

In summary,  we describe a sensitive \hii \ observation towards NGC1052-DF2 with the GMRT. We report a non-detection of atomic gas in the galaxy with $\textrm{M}_\textrm{HI} < 3.15 \times 10^6 \ ({\Delta V}/20 \ \textrm{km/s})^{1/2} \  \Ms \ (3\sigma)$ corresponding to very low gas fraction limit of $\gasfrac < 0.016 \ ({\Delta V}/20 \ \textrm{km/s})^{1/2} \ (3\sigma)$. The non-detection does not lead to a  resolution of the debate around NGC1052-DF2 containing little dark matter but suggests that the galaxy is similar to extremely gas-poor dwarf ellipticals; thus it is unlikely that a deeper observation to detect the \hii \ from this galaxy can be done using realistic telescope times. This study also suggests that a broad survey of atomic gas in UDGs residing in dense environments will be important to study their formation mechanism and evolutionary path viz. their lower density counterparts.  

\section*{Acknowledgements}
We thank the staff of the GMRT who have made these observations possible. GMRT is run by the National Centre for Radio Astrophysics of the Tata Institute of Fundamental Research. It is a great pleasure to thank Jayaram Chengalur and Nissim Kanekar for many useful discussions on the topic. I would also like to thank Andre Offringa for his prompt responses to queries regarding the usage of AOFLagger.




\bibliographystyle{mnras}
\bibliography{bibliography} 

\bsp	
\label{lastpage}
\end{document}